\documentclass[preprint,showpacs,preprintnumbers,amsmath,amssymb]{revtex4}

\usepackage{graphicx}
\usepackage{dcolumn}
\usepackage{bm}

\begin{document}

\preprint{APS/123-QED}

\title{On the nuclear stopping in asymmetric colliding nuclei   
\\}

\author{Varinderjit Kaur}
\author{Suneel Kumar}%
\email{suneel.kumar@thapar.edu}

\affiliation{%
School of Physics and Materials Science, Thapar University Patiala-147004, Punjab (India)\\
}
\author{Rajeev K. Puri}%
\affiliation{
Department of Physics, Panjab University, Chandigarh (India)\\
}%

\date{\today}

\begin{abstract}
Using an isospin-dependent quantum molecular dynamics (IQMD) model, nuclear stopping is analyzed in 
asymmetric colliding channels by keeping the total mass fixed. The calculations 
have been carried by varying the asymmetry
of the colliding pairs with different neutron-proton ratios in center of mass energy 250 MeV/nucleon 
and by switching off the effect of Coulomb interactions. 
We find sizable effect of asymmetry of colliding pairs on the stopping and therefore on the
equilibrium reached in a reaction.

\end{abstract}
\maketitle
\baselineskip=18pt
\section{Introduction}

In recent years, the study of heavy-ion collisions from low to relativistic energies 
has been focused on variety of phenomena that includes the multi-fragmentation \cite{multi}, anisotropy 
in the momentum distribution \cite{flow,Aichelin}, as well as global stopping of the nuclear matter 
\cite{1988}. 
 As we know, nuclear stopping is one of the essential observables that depends crucially on the reaction 
dynamics. The stopping has also been linked with the degree of thermalization and equilibrium 
reached in a reaction. This problem has been handled both theoretically and experimentally 
in recent years. Some have also tried to correlate it with the production of light 
charged particles. Even the role of symmetry energy has also been explored in this domain. It still
remains to be seen how stopping (and/or thermalization/equilibrium) is affected when the asymmetry of the
reacting partners is altered. As noted, the dynamics and energy deposition in asymmetric reaction can be 
quite different than in a symmetric reaction. The asymmetry of a reaction has also been reported to affect 
the collective flow, balance energy as well as elliptical flow of the colliding pairs. It is worth 
mentioning that the outcome and physical mechanism behind the symmetric and asymmetric reactions 
are entirely different.\\
Following the establishment of radioactive beam facilities in many laboratories, it became possible to
study the neutron-rich (or proton-rich) nuclear collisions at intermediate energies. 
The idea of studying nuclear stopping phenomena was introduced by Bass {\it et al.}, \cite{bass} via the
'isospin-mixing' method. In 1988, Bauer \cite{1988}, pointed out
that the nuclear stopping at intermediate energies is determined by the mean-field as well as by the
in-medium NN cross sections. In 1998, Li {\it et al.}, \cite{1998} found that the degree of isospin
equilibrium depends sensitively on both the in-medium NN cross section and equation of state 
of asymmetric nuclear matter. 
Another study in 2001 \cite{2001}, explored the possibility of using nuclear
stopping to probe the isospin dependence of in-medium NN cross-section. In 2002 \cite{2002}, authors 
studied the behavior of excitation function ${Q_{zz}/nucleon}$ and concluded that ${Q_{zz}/nucleon}$ can 
provide information about the isospin dependence in terms of binary cross-sections. The recent 
work of many authors suggested \cite{1998,2001,2002}  
that the degree of approaching isospin equilibration helps in probing the nuclear stopping 
in HIC. Several more studies also
focused in recent years on the isospin degree of freedom \cite{2004-08}.   
In 2006, one of us and co-workers \cite{jkdhawan} correlated the multifragmentation with global nuclear
stopping. Their study revealed that the light charged particles (LCP's) act in a similar 
manner as the anisotropy ratio. They, however, excluded the isospin content of the colliding nuclei.
In a recent communication \cite{2010}, one of us and co-workers tried to study the effect of 
symmetry energy and isospin-dependent cross-section
on nuclear stopping. Our findings revealed that the degree of stopping depends weakly on the
symmetry energy and strongly on the isospin-dependent cross-section. Most of the above 
mentioned theoretical and experimental studies concentrated on the dynamics for symmetric reacting 
partners. \\ 
It, however, remains to be seen how asymmetry of a colliding pair affects the stopping and thermalization.
As noted by FOPI collaboration \cite{FOPI}, asymmetry of colliding pairs \cite{asymm} can play decisive 
role in reaction dynamics. 
We plan to address this question in this present paper. In order to focus exclusively on the
asymmetric aspects, we shall vary the masses of the projectile and target in such a manner that total 
system mass remains constant. Another hidden parameter that can affect outcome in asymmetric colliding
nuclei is incident energy, we shall also keep center of mass energy fixed in present study.
This study is conducted within the framework of IQMD model which is explained in the Sec.-II. The results 
are presented in Sec.-III, leading to the conclusions in Sec.-IV.\\ 
\section{The Model}
The isospin-dependent quantum molecular dynamics (IQMD)\cite{hartnack} model treats different 
charge states of nucleons, deltas and pions explicitly, as inherited from the 
Vlasov-Uehling-Uhlenbeck (VUU) model. The IQMD model was used successfully in analyzing the large number 
of observables from low to relativistic energies. The isospin degree 
of freedom enters into the calculations via both cross sections and mean field. \\
In this model, baryons are represented by Gaussian-shaped density distributions \\
\begin{equation}
f_i(r,p,t)= \frac{1}{{\pi}^2{\hbar}^2}e^{\frac{{-(r-r_i(t))^2}}{2L}}e^{\frac{{-(p-p_i(t))^2}.2L}{\hbar^2}}.
\end{equation}
Nucleons are initialized in a sphere with radius R = ${1.12A^{1/3}}$ fm, in accordance with the  liquid 
drop model. The Gaussian width which we are taking is in close agreement as given in ref.  
\cite{gaussian}. Each nucleon occupies a volume of ${\hbar^3}$ so that phase space is uniformly filled.
The initial momenta are randomly chosen between 0 and Fermi momentum ${p_F}$. The nucleons of the target 
and projectile interact via two and three-body Skyrme forces and Yukawa potential. The isospin degrees of 
freedom is treated explicitly by employing a symmetry potential and explicit Coulomb forces between 
protons of the colliding target and projectile. This helps in achieving the correct
distribution of protons and neutrons within the nucleus.\\
The hadrons propagate using Hamilton equations of motion:\\
\begin{equation}
\frac{d\vec{r_i}}{dt} = \frac{d<H>}{d\vec{p_i}}~~~~;~~~~\frac{d\vec{p_i}}{dt} = -\frac{d<H>}{d\vec{r_i}}.
\end{equation}
 with\\ 
${ <H> = <T> + <V>}$  is the Hamiltonian.
\begin{equation}
    =  \sum_i\frac{p_i^2}{2m_i} + \sum_i \sum_{j>i}\int f_i(\vec{r},\vec{p},t)V^{ij}(\vec{r'},\vec{r}) 
f_j(\vec{r'},\vec{p'},t)d\vec{r}d\vec{r'}d\vec{p}d\vec{p'}.
\end{equation}
The baryon-baryon potential ${V^{ij}}$, in the above relation, reads as\\
\begin{eqnarray}
V^{ij}(\vec{r'}-\vec{r}) &~=~& V_{Skyrme}^{ij} + V_{Yukawa}^{ij} + V_{Coul}^{ij} + V_{Sym}^{ij}\nonumber\\
&=&t_1\delta(\vec{r'}-\vec{r})+ t_2\delta(\vec{r'}-\vec{r}){\rho}^{\gamma-1}(\frac{\vec{r'}+\vec{r}}{2})\nonumber\\ 
&+& t_3\frac{exp({\mid{\vec{r'}-\vec{r}}\mid}/{\mu})}{({\mid{\vec{r'}-\vec{r}}\mid}/{\mu})} + \frac{Z_{i}Z_{j}{e^2}}{\mid{\vec{r'}-\vec{r}\mid}}\nonumber\\
&+& t_{4} \frac{1}{\rho_o}T_{3}^{i}T_{3}^{j}.\delta(\vec{r'_i} - \vec{r_j}).
\end{eqnarray}
Where ${{\mu} = 1.5 fm}$, ${t_3 = -6.66 MeV}$, ${t_4 = 100 MeV}$. The values of ${t_1}$ and ${t_2}$ 
depends on the values of ${\alpha}$, ${\beta}$, and ${\gamma}$ \cite{Aichelin}.
Here ${Z_i}$ and ${Z_j}$ denote the charges of the ${i^{th}}$ and ${j^{th}}$ baryon, and ${T_{3}^i}$, 
${T_{3}^j}$ are their respective ${T_3}$ components (i.e. 1/2 for protons and -1/2 for neutrons). 
The parameters ${\mu}$ and ${t_1,........,t_6}$ are adjusted to the real part of
the nucleonic optical potential. \\


\section{Results and Discussion}

In the present study, projectile mass is varied between 16 and 56 units and targets are chosen as different
 isotopes of ${Xe}$, ${Sn}$, ${Ru}$ in such a way that total mass of the reaction remains constant 
(= 152) for all 
channels. For example, we take the reactions of  
$_{8}O^{16}+_{54}Xe^{136}$, $_{14}Si^{28}+_{54}Xe^{124}$, $_{16}S^{32}+_{50}Sn^{120}$, 
$_{20}Ca^{40}+_{50}Sn^{112}$, $_{24}Cr^{50}+_{44}Ru^{102}$, and $_{26}Fe^{56}+_{44}Ru^{96}$ etc. 
Although, the total mass remains constant, the asymmetry of the reaction 
$\eta = {\mid{(A_T-A_P)}/{(A_T+A_P)\mid}}$ keeps varying between 0.2 and 0.7.
Nuclear stopping in HIC has been studied with the help of different variables. A direct 
measure of nuclear stopping is the rapidity distribution defined as \cite{sanjeev_stopping}\\
\begin{equation}
Y(i) = \frac{1}{2}ln\frac{E(i)+p_{\|}(i)}{E(i)-p_{\|}(i)},
\end{equation}
where ${E(i)}$ and ${p_z(i)}$ are respectively, the energy and longitudinal momentum of the ${i^{th}}$ 
particle. For a complete stopping, one expects a single peaked Gaussian. Obviously, narrow Gaussian
indicates better thermalization (equilibrium) compared to broader ones.\\
Another quantity used in the literature \cite{sanjeev_stopping} is anisotropy ratio (R) defined as\\
\begin{equation}
R = \frac{2}{\pi}\frac{[\sum_i\mid{p_{\bot}(i)}\mid]}{[\sum_i\mid{p_{\|}(i)}\mid]} 
\end{equation}
where summation runs over all nucleons. The transverse ${p_{\bot}(i)}$ and longitudnal ${p_{\|}(i)}$
momenta reads respectively as ${\sqrt{p_x^{2}(i)+p_y^{2}(i)}}$ and  ${p_z(i)}$. Naturally, for a complete 
stopping R should be close to unity. Some studies use quadrupole moment ${Q_{zz}}$ to analyze the
stopping and thermalization. Quadrupole moment ${Q_{zz}}$ is defined as
\begin{equation}
Q_{zz} = \sum_i{[2p_{\|}^{2}(i)-p_{\bot}^{2}(i)]}
\end{equation}
Naturally, for a complete stopping, ${Q_{zz}}$ should be close to zero.\\

\begin{figure}
\includegraphics{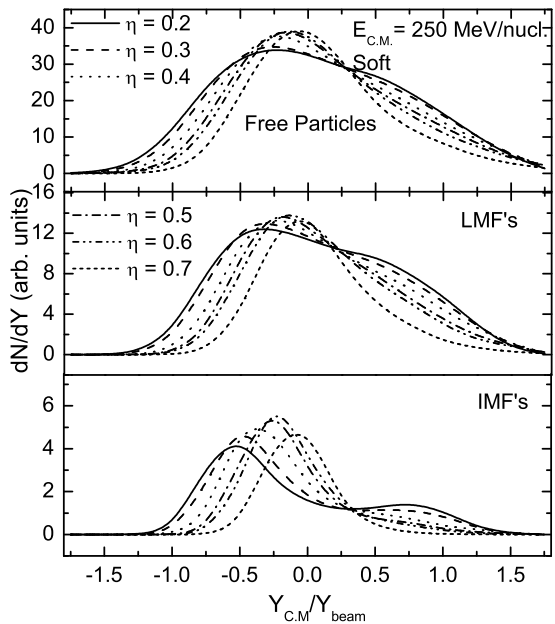}
\caption{\label{fig:1}  The rapidity distribution dN/dY as a function of reduced rapidity for free 
nucleons, LMF's and IMF's respectively at the center of mass energy ${E_{C.M.}}$ = 250 MeV/nucleon.
Different curves correspond to different asymmetries varying from 0.2 to 0.7 for semi-central
impact parameter.}  
\end{figure}
   
Fig.1 shows the rapidity distribution ${\frac{dN}{dY}}$ for the 
emission of free nucleons as well as light mass fragments (LMF's) [${(2\le A \le4)}$], and 
intermediate mass fragments (IMF's)[${(5\le A \le{A_{tot}}/6)}$] at a fixed centre-of-mass energy.
Note that the asymmetry varies between 0.2 and 0.7 corresponding to the reactions of 
$_{8}O^{16}+_{54}Xe^{136}$ (${\eta = 0.7}$), $_{14}Si^{28}+_{54}Xe^{124}$ (${\eta = 0.6}$), 
$_{16}S^{32}+_{50}Sn^{120}$ (${\eta = 0.5}$), $_{20}Ca^{40}+_{50}Sn^{112}$ (${\eta = 0.4}$),
$_{24}Cr^{50}+_{44}Ru^{102}$ (${\eta = 0.3}$), $_{26}Fe^{56}+_{44}Ru^{96}$ (${\eta = 0.2}$).
The ${Y_{C.M.}/Y_{beam} = 0}$ corresponds to mid-rapidity (participant) zone and hence is responsible 
for the hot 
and compressed zone. On the other hand, ${Y_{C.M.}/Y_{beam}}$ ${\ne}$ 0  corresponds to spectator zone  
(${Y_{C.M.}/Y_{beam} < -1}$ corresponds to target like (TL) 
and ${Y_{C.M.}/Y_{beam} > 1}$ corresponds to projectile like (PL) distributions). 
Interestingly, reaction corresponding to ${\eta = 0.7}$ yields peak at mid-rapidity which shifts
towards negative side when one considers nearly symmetric reactions. We see that majority of free 
particles and LMF's are emitted from the mid-rapidity region, whereas IMF's are emitted from the spectator 
matter. From the shape of the Gaussian, one sees that free particles and LMF's are  
better indicator for the thermal source at ${\eta}$ = 0.7.     \\ 
If these reaction channels are analyzed at a fixed lab energy, the situation would have been entirely 
different. In that case, nearly symmetric reactions ${\eta = 0.2}$ would yield better thermalized source.
It is worth mentioning that in many studies, one has 
kept the lab energy fixed. The absolute conclusion in these studies are not due to asymmetry alone.\\ 
     
In Fig.2, we display the impact parameter dependence of global variables (R and ${Q_{zz}}$) along with the
multiplicity of LMF's. The results are displayed at ${E_{C.M.}}$ = 250 MeV/nucleon.  
Different curves in each panal represent different asymmetries. We observe that
R and ${1/Q_{zz}}$ behave in a similar fashion (note that R and ${Q_{zz}}$ will behave in just 
opposite fashion). The amount of stopping/equilibrium increases with the asymmetry of the reaction.  
As we know, major contribution for the stopping of nuclear matter comes from the hot and compressed 
matter that decreases almost linearly with impact parameter. To correlate the degree of stopping with the 
multiplicity of light charged particles, we show in the last panal the impact parameter dependence of the 
multiplicity of light charged particles. 
\begin{figure}
\includegraphics{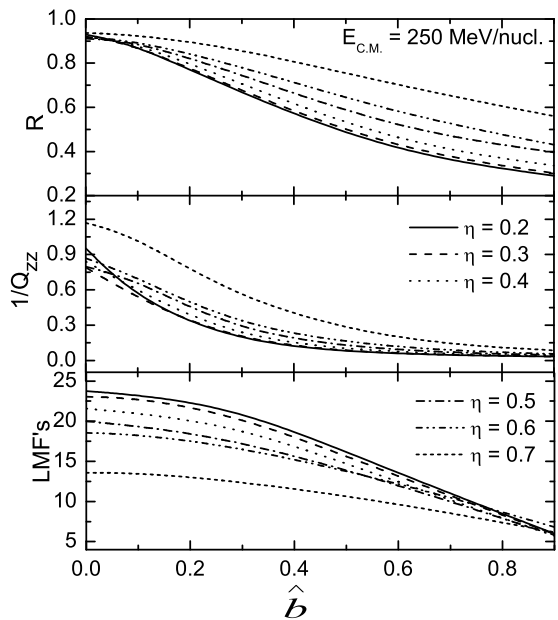}
\caption{\label{fig:2}  The anisotropy ratio R, quadrupole moment ${1/Q_{zz}}$ and multiplicity of
LMF's as a function of impact parameters. Different curves correspond to different asymmetries 
varying from 0.2 to 0.7 for semi-central impact parameters. }  
\end{figure}
The behavior of light charged particle with impact parameter is similar to that of anisotropic 
ratio and inverse of quadrupole moment. The decrease in the multiplicity of
LMF's complements with corresponding increase in the heavy fragments. These fragments
are the remanant of the spectator matter. Therefore, a decrease in the multiplicity of LMF's with 
impact parameter measures directly the decrease in the degree of equilibrium and hence 
global stopping. This implies that the LMF's production can act as an indicator for nuclear 
stopping. At the same time, we also note that behavior with respect to asymmetry does behave in same 
fashion. This happens due to the fact that participant zone increases when one moves towards 
nearly symmetric reactions. Therefore, more LMF's are produced for ${\eta}$ = 0.2 compared to ${\eta}$ = 
0.7. It is worth mentioning that in the case of multiplicity of LMF's, we talk about the number of 
particles while in the case of stopping, it is the momentum phase space (x, y and z components of momentum)
that is responsible for the growth.\\

The stopping at any time during the collision can be divided into the contributions emerging from the 
protons and neutrons. We decompose the stopping into the contributions due to protons
and neutrons. Here, at each time step during the collision, stopping due to neutrons and protons
is analyzed separately.\\ 
Fig.3(a) shows the final state quadrupole moment ${1/Q_{zz}}$ decomposed into contributions due to neutrons
and protons as a function of asymmetry ${\eta}$.
\begin{figure}
\includegraphics{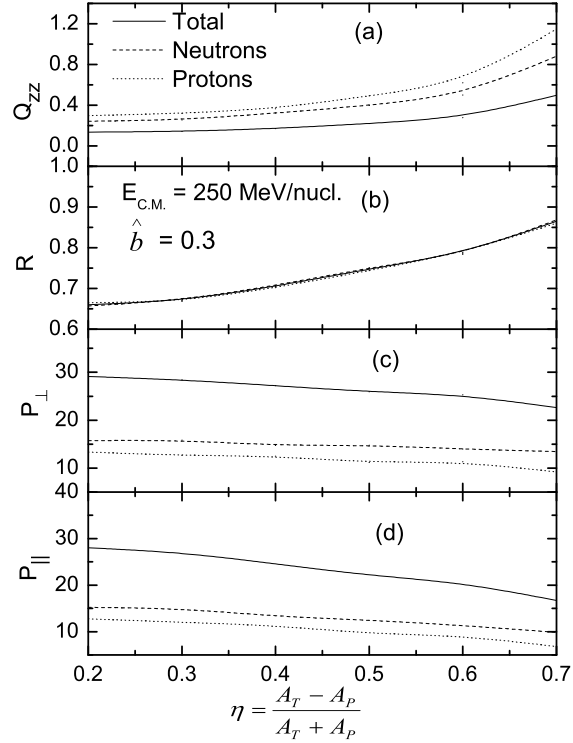}
\caption{\label{fig:3}  The quadrupole moment ${1/Q_{zz}}$ and anisotropy ratio R as a function of 
asymmetry ${\eta}$ at a center of mass energy ${E_{C.M.}}$ = 250 MeV/nucleon. The third and fourth
panals show the variation of transverse and longitudnal momentum with asymmetry, respectively. Different
curves correspond to the contribution of neutrons and protons along with total contribution. }  
\end{figure}
Fig.3(b) shows the final state anisotropy ratio ${<R>}$ as a function of the asymmetry of the system. 
We see no difference between the contributions due to neutrons and protons. This is due to the fact 
that ${<R>}$ is the ratio of the mean transverse momentum
${p_{\bot}(i)}$ to the mean longitudnal momentum  ${p_{\|}(i) = p_z(i)}$.
To see the clear contribution of neutrons and protons, one has to look into the contributions of transverse
and longitudnal momenta as shown in Fig.3(c) and 3(d), respectively. A linear enhancement in 
the transverse and longitudnal momentum can be seen with asymmetry ${\eta}$. Further the contribution of 
neutrons exceeds the corresponding contributions due protons.\\

It will be of further interest to see whether the above findings depend on the isospin asymmetry
(N/Z dependence) or not. For this, we display in Fig.4, the anisotropy ratio ${<R>}$, inverse of 
quadrupole moment ${Q{zz}}$ and multiplicity of LMF's as a function of neutron to proton ratio (N/Z) 
at different impact parameters ranging from central to peripheral one. We are also showing the
results at impact parameter $\hat{b}$ = 0.1 by changing the Gaussian width according to the formulae 
${\sigma = 0.16N^{1/3} + 0.49}$ as given in ref \cite{gaussian}. Interestingly, we see no change in the
results by the variation of Gaussian width.  
An increase in the number of neutrons 
will increase the number of collisions and hence the absolute value of ${<R>}$, ${1/Q_{zz}}$ will 
increase with N/Z ratio.
\begin{figure}
\includegraphics{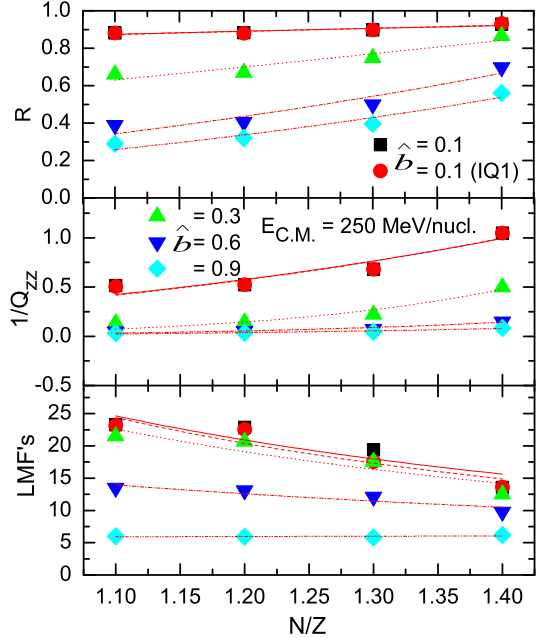}
\caption{\label{fig:4}(Color online) Same as Fig. 2, but as a function of N/Z ratio. Different curves 
correspond to different impact parameters ranging between central to peripheral one.}  
\end{figure}
Our exclusive findings are: (1) maximum stopping is obtained for the systems with larger neutron
content. This is true at all colliding geometries. This dependence diminishes as one moves from central to 
peripheral geometry. It is due to the fact that nuclear stopping 
is governed by the participant zone only. Moreover, in systems with more neutron content, role of 
symmetry energy could be larger, whereas effects due to isospin-dependent cross-section could play a 
dominant role in systems with less
neutron content (proton content). Therefore there is a possibility that the combined effect of symmetry
energy and cross-section could be approximately the same for all systems with different neutron and
proton content \cite{sakshi}. Also neutron-neutron or proton-proton 
cross-section is a factor of 3 lower than the neutron-proton cross-section. (2) On the other hand, the 
multiplicity of LMF's follows the 
reverse trend. This is due to the reason that in case of N/Z = 1.1 (${\eta}$ = 0.2), 
participant zone is more compared to N/Z = 1.4 (${\eta}$ = 0.7). Therefore, more LMF's are 
produced for N/Z = 1.1 as compared to N/Z = 1.4.       
\section{Conclusion}
Using the isospin-dependent quantum molecular dynamics (IQMD), we have studied the nuclear stopping 
in asymmetric colliding channels by keeping total mass fixed. The calculations have been carried 
out by varying 
the asymmetry of the colliding pairs with different neutron-proton ratios in the center of mass energy 
250 MeV/nucleon and by switching off the effect of Coulomb 
interactions. The contribution of the
neutrons and protons is checked in terms of anisotropy ratio ${<R>}$ and quadrupole moment ${Q_{zz}}$.
The maximum stopping is obtained for the systems having maximum
neutron-proton ratio because the contribution of neutrons remains enhanced throughout the 
asymmetry range. Moreover, this dependence becomes weaker as as one
moves from central to peripheral geometry. Interestingly, reverse trend is obtained when we vary 
the multiplicity of LMF's with N/Z ratio.


\section {Acknowledgment}
This work has been supported by the grant from Department of Science and Technology (DST) Government of 
India vide Grant No.SR/WOS-A/PS-10/2008.\\ 

\end{document}